\documentclass{ifacconf}
\usepackage{graphicx}      
\usepackage{natbib}        
\usepackage{amsmath}
\usepackage{amsfonts}
\usepackage{mathrsfs}
\let\theoremstyle\relax
\usepackage{amsthm}
\usepackage{booktabs}
\usepackage{multirow}
\usepackage{tabularx}
\usepackage{mathtools}
\newtheorem{lemma}{Lemma}
\newtheorem{proposition}{Proposition}

\newtheorem{assumption}{Assumption}
\theoremstyle{definition}

\newtheoremstyle{myremark}
  {3pt}   
  {3pt}   
  {\normalfont} 
  {}      
  {\bfseries} 
  {.}     
  {.5em}  
  {}
\theoremstyle{myremark}
\newtheorem{remark}{Remark}

\newcommand{\hop}{\mathsf{H}}
\usepackage{etoolbox}

\usepackage{comment}
\usepackage{xcolor}
\usepackage{tikz}
\usetikzlibrary{
    positioning,
    arrows.meta,
    shapes.geometric,
    calc,
    fit,
    decorations.pathmorphing
}
\usepackage{algorithm}
\usepackage{algpseudocode}
\usepackage{bbm}

\usepackage{eso-pic}
\AddToShipoutPictureBG*{%
\AtPageUpperLeft{%
\setlength\unitlength{1in}%
\hspace*{\dimexpr0.5\paperwidth\relax}
\makebox(0,-1.75)[c]{
\begin{tabular}{c c}
Enrico Dozzi \emph{et al.},
Event-Triggered Data Selection for Online Frequency-Domain Identification, \\
Manuscript submitted for publication, 
uploaded to ArXiv \today \\
\end{tabular}}}}

\begin{document}
\begin{frontmatter}

\title{Event-Triggered Data Selection for Online Frequency-Domain Identification}
\author[First]{Enrico Dozzi}, 
\author[First]{Tom Oomen},
\author[First]{Rodrigo A. González}

\address[First]{Department of Mechanical Engineering, Eindhoven University of Technology, Eindhoven, The Netherlands.}

\begin{abstract}
Online identification of frequency response functions is increasingly relevant for control and monitoring of high-performance systems whose dynamics may change during operation.
In modern machines, however, continuously transmitting all locally acquired measurements to a central unit for identification can impose substantial communication demands, particularly as sensing rates and the number of subsystems increase.
This paper develops an event-triggered data-selection approach for remote frequency-domain identification, in which periodically acquired measurements are evaluated locally and  transmitted only when relevant.
The selection mechanism combines information about the current estimation uncertainty with evidence of model mismatch, reducing communication while preserving identification accuracy and responsiveness to changes in the dynamics.
\end{abstract}    
\end{frontmatter}

\section{Introduction}
\vspace{-0.4cm}

Frequency response functions (FRFs) provide a widely used representation of system dynamics in mechatronic systems, with applications in controller design, performance assessment, and condition monitoring \citep{evers2020frequency, allemang2022frequency}.
FRFs characterize the input-output behavior of a system at the frequencies of interest, providing a physically interpretable description of its dynamics.
Frequency-domain system identification provides a well-established framework for estimating such models from measured input-output data \citep{pintelon2012system}.
Classical FRF descriptions typically rely on the assumption that the underlying system remain time invariant over the observation interval.

However, in many applications, dynamics cannot be regarded as fixed over the entire operating lifetime of the system.
Changes in operating conditions, gradual degradation, or abrupt faults may alter relevant dynamic properties \citep{pintelon2015nonparametric, classens2022fault}.
Thus, an FRF identified offline may cease to represent the current dynamics.
This motivates online frequency-domain identification, where the FRF estimate is continuously adapted as new measurements become available .
In modern high-performance machines, this online tracking must typically be performed across numerous interconnected and locally sensed subsystems, whose dynamics must be characterized to support system-level monitoring and control.

High sensing rates across these distributed subsystems generate rapidly growing streams of input–output data available for identification.
If every locally acquired measurement is transmitted to a central unit and used for model updates, the associated communication and computational requirements grow with both the sampling rate and the number of subsystems, becoming a limiting factor as system complexity increases \citep{ge2019distributed}. 
At the same time, consecutive high-rate measurements may not contribute equally to the quality of the identified model: once the dynamics are well characterized, many measurements may provide only limited additional information.
This motivates compressing the measurement stream according to its relevance for identification, such that each local subsystem should communicate only those measurements that contribute sufficient information about its dynamics.

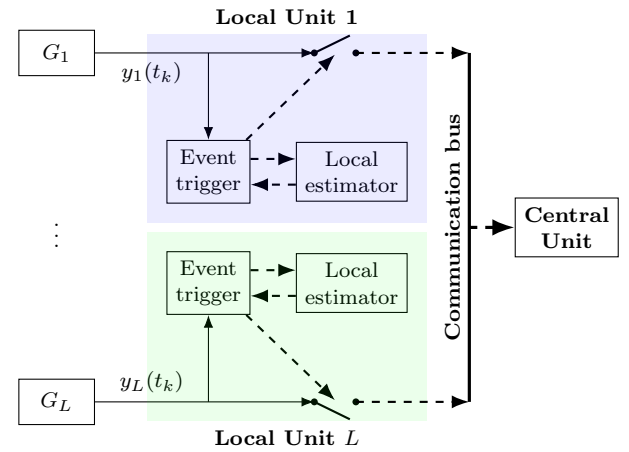
\begin{figure}
\centering
\begin{tikzpicture}[
    >=Latex,
    font=\small,
    block/.style={
        draw,
        rectangle,
        minimum width=1cm,
        minimum height=0.6cm,
        align=center
    },
    local/.style={
        draw,
        dashed,
        rectangle,
        inner sep=0.25cm
    }
]


\node[block] (plant1) {$G_1$};

\coordinate (branch1) at ([xshift=1.5cm]plant1.east);
\coordinate (swL1)    at ([xshift=2.9cm]plant1.east);
\coordinate (swR1)    at ([xshift=0.55cm]swL1);

\node[
    block,
    below=1.15cm of branch1,
    xshift=0cm,
    minimum width=1cm
] (trigger1)
{Event\\trigger};

\node[
    block,
    right=0.6cm of trigger1,
    minimum width=1cm
] (estimator1)
{Local\\estimator};

\draw[->]
(plant1.east) --
node[below, xshift=-0.7cm] {$y_1(t_k)$}
(swL1);

\draw[->]
(branch1) --
(trigger1.north);

\fill (swL1) circle (1.3pt);
\fill (swR1) circle (1.3pt);

\draw[thick]
(swL1) --
([xshift=-0.08cm,yshift=0.23cm]swR1);

\coordinate (switchcontrol1)
    at ($(swL1)!0.5!(swR1)+(0,-0.02)$);

\draw[->, dashed, thick]
([xshift=0.5cm]trigger1.north) --
(switchcontrol1);

\draw[->, dashed, thick]
([yshift=1.7mm]trigger1.east) --
([yshift=1.7mm]estimator1.west);

\draw[->, dashed, thick]
([yshift=-1.7mm]estimator1.west) --
([yshift=-1.7mm]trigger1.east);


\node[
    block,
    below=4cm of plant1
] (plantL) {$G_L$};

\coordinate (branchL) at ([xshift=1.5cm]plantL.east);
\coordinate (swLL)    at ([xshift=2.9cm]plantL.east);
\coordinate (swRL)    at ([xshift=0.55cm]swLL);

\node[
    block,
    above=1.15cm of branchL,
    xshift=0cm,
    minimum width=1cm
] (triggerL)
{Event\\trigger};

\node[
    block,
    right=0.6cm of triggerL,
    minimum width=1cm
] (estimatorL)
{Local\\estimator};

\draw[->]
(plantL.east) --
node[above, xshift=-0.7cm] {$y_L(t_k)$}
(swLL);

\draw[->]
(branchL) --
(triggerL.south);

\fill (swLL) circle (1.3pt);
\fill (swRL) circle (1.3pt);

\draw[thick]
(swLL) --
([xshift=-0.08cm,yshift=-0.23cm]swRL);

\coordinate (switchcontrolL)
    at ($(swLL)!0.5!(swRL)+(0,0.02)$);

\draw[->, dashed, thick]
([xshift=0.5cm]triggerL.south) --
(switchcontrolL);

\draw[->, dashed, thick]
([yshift=1.7mm]triggerL.east) --
([yshift=1.7mm]estimatorL.west);

\draw[->, dashed, thick]
([yshift=-1.7mm]estimatorL.west) --
([yshift=-1.7mm]triggerL.east);


\node[
    local,
    draw=none,
    fill=blue,
    fill opacity=0.08,
    fit=(branch1)(swL1)(swR1)(trigger1)(estimator1),
    label=above:{\textbf{Local Unit 1}}
] (local1) {};

\node[
    local,
    draw=none,
    fill=green,
    fill opacity=0.08,
    fit=(branchL)(swLL)(swRL)(triggerL)(estimatorL),
    label=below:{\textbf{Local Unit $L$}}
] (localL) {};


\node at ($(plant1)!0.5!(plantL)$) {$\vdots$};


\coordinate (busTop)
    at ([xshift=1.5cm]swR1);

\coordinate (busBottom)
    at ([xshift=1.5cm]swRL);

\draw[very thick]
(busTop) -- (busBottom);

\node[
    rotate=90,
    anchor=south
] at ($(busTop)!0.5!(busBottom)$)
{\textbf{Communication bus}};


\draw[->, dashed, thick]
(swR1) --
(swR1 -| busTop);

\draw[->, dashed, thick]
(swRL) --
(swRL -| busTop);


\node[
    block,
    right=0.6cm of $(busTop)!0.5!(busBottom)$,
    minimum width=1cm,
    minimum height=0.6cm
] (central)
{\textbf{Central} \\ \textbf{Unit}};

\draw[->, very thick, dashed]
($(busTop)!0.5!(busBottom)$) --
(central.west);

\end{tikzpicture}

\caption{Distributed event-triggered online frequency-domain identification architecture with $L$ subsystems.}
\label{fig:distributed_framework}
\end{figure}

A natural approach is event-triggered data selection, in which each local subsystem acquires measurements at a prescribed rate and transmits to a central unit only those deemed sufficiently informative for remote identification, as shown in Figure \ref{fig:distributed_framework}.
The objective is to reduce the amount of transmitted data while preserving the information required for accurate online identification and timely adaptation to changes in the system dynamics.
A key challenge introduced by the event-triggered selection mechanism is that measurements used for identification form an irregular and data-dependent sequence, disrupting the regular temporal structure exploited by standard frequency-domain approaches \citep{pintelon2012system}.

\subsection{Contribution}
The main contribution of this work is an event-triggered data-selection framework that determines online which locally acquired measurements are sufficiently informative to be transmitted for remote frequency-domain identification.
The event trigger evaluates each candidate measurement according to two complementary criteria: whether it can contribute new information about the system dynamics, and whether it reveals a discrepancy between the observed output and the current model prediction.
To support this decision, the local sensing unit maintains a recursive frequency-domain estimator that summarizes the information contained in previously selected measurements through the current FRF estimate and its associated uncertainty.
Under multisine excitation, this estimator directly estimates the FRF at the excited frequencies from time-domain measurements and naturally accommodates the irregular sequence generated by event-based data selection.
Measurements selected by the trigger are transmitted to the central unit, where they form a sparse, information-driven dataset for online identification.
The resulting framework reduces communication while retaining a sufficiently informative dataset for accurate frequency-domain identification and timely adaptation to abrupt changes in the system dynamics.

\subsection{Related Work}
Event-based mechanisms have been extensively studied to reduce communication in related areas, including control \citep{heemels2012introduction,miskowicz2018event,postoyan2025event}, and state estimation \citep{petri2021event,petri2024decentralized}.
The setting considered here shares the architecture of periodic event-triggered control \citep{heemels2012periodic}: measurements are acquired periodically and the triggering condition is evaluated only at the sampling instants, which guarantees a strictly positive inter-event time.
In the present work, however, the triggering mechanism is designed to select informative data for system identification rather than to enforce closed-loop control objectives.
A related triggering principle appears in the variance-based state estimation approach of \citet{trimpe2014event}, where communication decisions depend on the uncertainty associated with the predicted measurement.
Here, uncertainty is instead used to assess how much additional information for identification each measurement can provide.
Moreover, the triggering decision is not based on uncertainty alone, but also on the discrepancy between the newly observed measurement and the current prediction, making the data-selection mechanism reactive to changes.

Related ideas have also been considered in system identification under communication constraints.
Measurement-scheduling approaches determine which sensor observations are transmitted to a remote estimator balancing estimation accuracy and communication requirements \citep{you2013asymptotically,han2015optimal}.
Event-triggered identification has also been studied for specific model and measurement structures, including FIR systems with binary-valued observations \citep{diao2018event} and distributed identification of linear stochastic systems \citep{geng2025distributed}.
These works consider systems with fixed dynamics and estimate finite-dimensional parameters of time-domain models.
In contrast, the present work addresses online nonparametric frequency-domain identification of systems whose dynamics may change abruptly over time.

A distinct line of work considers event-based measurement acquisition for system identification, relying on sensing mechanisms that generate samples when prescribed events occur \citep{kawaguchi2016system,gonzalez2023algorithm,gonzalez2024kernel}.
The present work instead targets systems equipped with conventional periodic sensing, with the event-triggered mechanism acting on data selection rather than on physical measurement acquisition.
This makes the proposed approach applicable without requiring event-based sampling hardware.
The present work is also related to our previous study \citep{dozzi2026d}, where the sampling pattern was designed offline to maximize the expected information available for identification.
Here, the information-based selection mechanism is brought online: candidate measurements are evaluated sequentially according to the information already retained from previously selected data.

\textit{Outline.}
The remainder of this work is organized as follows.
Section \ref{sec:problem_formulation} formulates the identification problem.
Section \ref{sec:event_triggered_identification} presents the proposed event-triggered data-selection framework.
Section \ref{sec:analysis} analyzes the information properties of the triggering rule.
Section \ref{sec:simulation_results} evaluates the proposed framework on a mass-spring-damper-mass system.
Section~\ref{sec:conclusions} summarizes the main findings.

\textit{Notation.}
The sets $\mathbb{Z}$, $\mathbb{R}$, $\mathbb{C}$ denote the sets of all integer, real and complex numbers, respectively.
Vectors are considered as column vectors, unless transposed.
The notation ${0}_n$ denotes the $n$-dimensional vector of zeros.
The identity matrix of dimension $n\times n$ is indicated by $I_n$.
The Euclidean norm of the vector $v$ is denoted by \( ||v||_2 \) and the spectral norm of the matrix $A$ is denoted by $||A||_2$.
Operator $\mathrm{p}$ denotes the differentiation operator \(\mathrm{p} := {\text{d}}/{\text{d}t}\). The complex conjugate and complex conjugate transpose of a vector $x$ are denoted as $x^*$ and $x^{\hop}$, respectively. 
The symbol $\mathbbm{1}_{\{\cdot\}}$ denotes the indicator function, equal to $1$ when the enclosed condition is true and $0$ otherwise.

\section{Problem Formulation}
\label{sec:problem_formulation} 

We consider a continuous-time, single-input single-output (SISO) system with input $u(t)\in\mathbb{R}$ and noise-free output $x(t)\in\mathbb{R}$, whose dynamics are described by $G(\mathrm p,t)$ and may undergo abrupt changes over time.

\begin{assumption}[Piecewise-LTI dynamics]
\label{ass:abrupt_fault}
There exists a sequence of unknown change instants $0=T_0<T_1<T_2<\cdots$ such that $G(\mathrm p,t)=G_j(\mathrm p)$ for $T_j\le t<T_{j+1}$, where each \(G_j(\mathrm p)\) is a deterministic LTI operator.
\end{assumption}

\begin{figure*}[t]
\centering
\begin{tikzpicture}[
    >=Latex,
    font=\small,
    node distance=1.2cm,
    block/.style={
        draw,
        rectangle,
        minimum width=1.5cm,
        minimum height=1.2cm,
        align=center
    }
]


\node[block, label=above:{\textbf{Plant}}] (sys)
{$G(\mathrm p,t)$};

\node[block, right=2cm of sys] (realsampler)
{Periodic\\sampler\\$t_k=kT_s$};

\node[block, draw=none, right=2.5cm of sys](sampler){};

\coordinate (swL) at ([xshift=1.5cm]sampler.east);
\coordinate (swR) at ([xshift=0.8cm]swL);

\node[block, right=2.5cm of swR] (remote)
{Central\\Unit};


\node[
    block,
    below=1.7cm of $(swL)!0.5!(swR)$,
    xshift=-5mm,
    minimum width=2cm
] (trigger)
{Event trigger\\
$\Gamma(\mathcal I(t_k),y(t_k))\geq\eta$};

\node[
    block,
    left=1.2cm of trigger,
    minimum width=2.0cm
] (estimator)
{Recursive\\estimator};


\draw[->]
([xshift=-1.5cm]sys.west)
--
node[above] {$u(t)$}
(sys.west);

\draw[->]
(sys.east)
--
node[above] {$x(t)$}
(realsampler.west);

\draw[->]
(realsampler.east)
--
node[above] {$y(t_k)$}
(swL);

\fill (swL) circle (1.4pt);
\fill (swR) circle (1.4pt);

\draw[thick]
(swL)
--
([xshift=-0.10cm,yshift=0.28cm]swR);

\draw[->, dashed, thick]
(swR)
--
node[above, align=center, xshift=3mm]
{$\tau_i$, $y(\tau_i)$}
(remote.west);


\draw[->]
([xshift=-0.5cm]swL)
--
([xshift=-0.4cm]trigger.north);

\coordinate (switchcontrol)
    at ($(swL)!0.5!(swR)+(0,-0.05)$);

\draw[->, dashed, thick]
([xshift=5mm]trigger.north)
--
node[right] {Event}
(switchcontrol);


\draw[->, dashed, thick]
([yshift=2mm]trigger.west)
--
node[above, align=center]
{$\tau_i$, $y(\tau_i)$}
([yshift=2mm]estimator.east);

\draw[->, dashed, thick]
([yshift=-2mm]estimator.east)
--
node[below, align=center]
{$\hat{\theta}_i$, $P_i$}
([yshift=-2mm]trigger.west);


\node[
    draw=none,
    fill=blue,
    fill opacity=0.08,
    rectangle,
    inner sep=0.3cm,
    fit=(sampler)(swL)(swR)(trigger)(estimator),
    label=above:{\textbf{Local Unit}}
] (local) {};

\end{tikzpicture}

\caption{Local event-triggering mechanism for measurement selection and remote frequency-domain identification.}
\label{fig:framework}
\end{figure*}
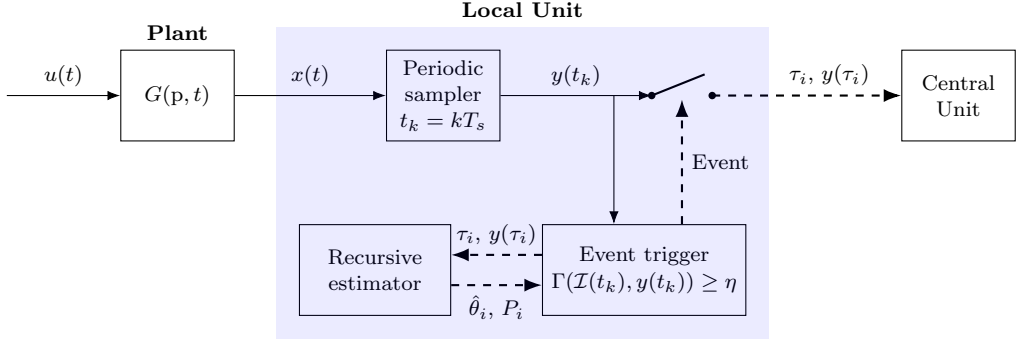

\begin{remark}
The theoretical development considers a single subsystem and its local selection mechanism, as shown in Figure~\ref{fig:framework}.
Since the triggering decision is performed locally, the same mechanism can be applied to multiple subsystems without additional technical difficulties.
\end{remark}

The system is excited by a multisine input signal 
\begin{equation}
u(t) = a_0+\sum_{m=1}^{M} a_m \cos(\omega_m t + \phi_m),
\label{eq:multisine}
\end{equation} 
where \(\{\omega_m\}_{m=1}^M\) are the chosen nonzero excitation frequencies, and \(a_m\), \(\phi_m\) are the corresponding amplitudes and phase shifts.
The resulting set of excited frequencies is \(\Omega:= \{0\}\cup\{\pm \omega_m\}_{m=1}^M\).
The input signal $u(t)$ is assumed to be continuously available.

The system output $x(t)$ is measured periodically at sampling instants $\{t_k\}_{k\ge 1}=\{kT_s\}_{k\ge 1}$ with sampling period $T_s>0$, producing the noisy measurements
\begin{equation}
    y(t_k)=x(t_k)+e(t_k),
    \label{eq:measured_output_y_tk}
\end{equation}
where $e(t_k) \in \mathbb{R}$ is additive measurement noise at time $t_k$.  

\begin{assumption}[Measurement noise]
\label{assump:noise}
The measurement noise sequence $\{e(t_k)\}_{k\ge 1}$ is independent and identically distributed Gaussian with zero mean and variance $\sigma^2$.
\end{assumption} 

The event-based measurement sequence is generated online by selecting a subset of the periodically acquired samples $\{y(t_k)\}_{k\geq1}$.
Starting from the most recent event instant $\tau_i$, each subsequent candidate measurement $y(t_k)$ is evaluated sequentially using the information currently available at the sensing unit.
The next event instant is
\begin{equation}
    \tau_{i+1}
    =
    \inf
    \left\{
    t_k>\tau_i:
    \Gamma(\mathcal I(t_k), y(t_k))\geq \eta
    \right\},
    \label{eq:general_trigger}
\end{equation}
where $\eta$ is a design threshold,  $\Gamma(\cdot)$ is a scalar triggering functional, and $\mathcal I(t_k)$ denotes the information available at the sensing unit immediately before the new measurement $y(t_k)$ is observed.
At an event instant $\tau_{i+1}$, the selected measurement $y(\tau_{i+1})$ is transmitted to the central unit and incorporated into the identification dataset, while measurements for which \eqref{eq:general_trigger} is not satisfied are omitted.
Transmitting selected measurements rather than only a locally computed estimate preserves flexibility at the central unit, where the dataset can be processed within a common identification framework across multiple subsystems. 
The central estimator is not required to replicate the local implementation, provided it is consistent with the same underlying model used to assess measurement relevance.

The objective is to compress the periodically acquired measurements $\{y(t_k)\}_{k\geq1}$ into an informative sequence $\{y(\tau_i)\}_{i\geq1}$ for remote frequency-domain identification.
To this end, we formulate a recursive frequency-domain estimator that summarizes the information contained in previously selected measurements, and design the triggering functional $\Gamma(\cdot)$ in \eqref{eq:general_trigger} to assess the relevance of each newly acquired measurement based on this information.
The resulting selection mechanism should allow accurate FRF estimation and timely adaptation to abrupt changes.

\section{Event-Triggered Data Selection for Online Frequency-Domain Identification}
\label{sec:event_triggered_identification}

This section develops the event-triggered data-selection mechanism.
At the local unit, a recursive frequency-domain estimator is maintained to summarize the information contained in previously selected measurements.
Recursive methods provide a natural framework for online identification by updating the model incrementally without requiring the storage of complete input-output datasets \citep{ljung1983theory}.
The estimator is used by the event-triggering mechanism to assess each newly acquired candidate measurement and determine whether it should be transmitted to the central unit.
Hence, we first formulate the recursive estimator and then use the information it maintains to construct the event-triggering rule.

\subsection{Recursive Frequency-Domain Estimator}
\label{subsec:recursive_estimator}

Under multisine excitation \eqref{eq:multisine}, the output measurements admit the complex-valued linear regression representation
\begin{equation}
    y(t_k)=\varphi^{\hop}(t_k)\theta(t_k)+\varepsilon(t_k), 
    \label{eq:linear_regression}
\end{equation}
following the frequency-domain formulation of \cite{gonzalez2025finite}.
Here, $\varepsilon(t_k)=y(t_k)-\varphi^{\hop}(t_k)\theta(t_k)$ denotes the regression error, collecting measurement noise and deviations from the steady-state frequency-domain representation, such as transient and time-varying effects. 
Furthermore, $\theta(t_k)\in\mathbb{C}^{2M+1}$ contains the FRF of the system at time $t_k$ and $\varphi(t_k)\in\mathbb{C}^{2M+1}$ is the regressor vector evaluated at time $t_k$, i.e.,
\begin{align}
    \theta(t_k) = 
    \begin{bmatrix}
        G(0, t_k) \\ G(j\omega_1, t_k) \\ 
        G(-j\omega_1, t_k) \\ \vdots \\
        G(j\omega_M, t_k) \\ G(-j\omega_M, t_k)
    \end{bmatrix}, \quad
        \varphi(t_k) =
    \begin{bmatrix}
        a_0 \\[2pt]
        \tfrac{a_1}{2} e^{-j(\omega_1 t_k+\phi_1)} \\
        \tfrac{a_1}{2} e^{j(\omega_1 t_k+\phi_1)} \\
        \vdots \\
        \tfrac{a_M}{2} e^{-j(\omega_M t_k+\phi_M)} \\
        \tfrac{a_M}{2} e^{j(\omega_M t_k+\phi_M)}
    \end{bmatrix}.
    \label{eq:varphi}
\end{align}
Only the measurements selected by the event-triggering mechanism \eqref{eq:general_trigger} are used to update the recursive estimator.
Thus, the identification algorithm operates on the event-generated regression equations $y(\tau_i) = \varphi^{\hop}(\tau_i)\theta(\tau_i) + \varepsilon(\tau_i)$, which are obtained by restricting \eqref{eq:linear_regression} to the selected measurements.
Under Assumption~\ref{ass:abrupt_fault}, the parameter vector is piecewise constant, i.e., $\theta(t)=\theta_j$ for $T_j\le t<T_{j+1}$, where $\theta_j$ contains the FRF of the LTI dynamics $G_j(p)$ at the excited frequencies.

To estimate the current FRF from the event-selected measurements while allowing adaptation after changes in the dynamics, we employ an exponentially weighted recursive least-squares (RLS) estimator.
Adapting the standard RLS \citep{ljung1983theory} to the complex-valued regression and event-triggered measurements gives
\begin{align}
    &L_N = \frac{P_{N-1}\varphi(\tau_N)}{\lambda+\varphi^{\hop}(\tau_N)P_{N-1}\varphi(\tau_N)} \nonumber \\
    &P_N = \frac{1}{\lambda} \left[ P_{N-1}-L_N\varphi^{\hop}(\tau_N)P_{N-1} \right] \nonumber \\
    &\hat{\theta}_N = \hat{\theta}_{N-1}+L_N\left[y(\tau_N)-\varphi^{\hop}(\tau_N)\hat{\theta}_{N-1}\right],
    \label{eq:RLS_multisine}
\end{align}
where $L_N\in\mathbb{C}^{2M+1}$ is the recursive gain vector determining the contribution of the newly selected measurement to the parameter update, $P_N\succ0$ provides a measure of the uncertainty associated with the current FRF estimate, and $\lambda\in(0,1]$ is the forgetting factor, which determines how rapidly past information is down-weighted as new measurements are incorporated into the estimator.

\begin{remark}
    Forgetting is indexed by the number of event-selected measurements rather than by elapsed time.
    Thus, past information is discounted only when a new measurement is selected, coupling forgetting to the arrival of informative data.
    The effective memory of the estimator in physical time adapts to the event rate: it is preserved when no event occurs, while an increased event rate following a change in the dynamics causes information associated with the previous regime to be replaced more rapidly.
\end{remark}

Defining the corresponding information matrix as $R_N=P_N^{-1}$, its evolution is given by
\begin{equation}
    R_N
    =
    \lambda^N R_0
    +
    \sum_{i=1}^N
    \lambda^{N-i}
    \varphi(\tau_i)\varphi^{\hop}(\tau_i),
    \label{eq:P_inverse}
\end{equation}
so that $R_N$ accumulates the exponentially weighted information provided by the event-selected measurements together with the initial information specified through $R_0=P_0^{-1}$.
The recursion is initialized with $\hat{\theta}_0={0}_{2M+1}$ and $P_0=\alpha I_{2M+1}$, where $\alpha>0$ is chosen sufficiently large to represent weak initial information about the FRF and make the initialization term in \eqref{eq:P_inverse} negligible.
Thus, the resulting recursion provides the local unit with a compact state $(\hat{\theta}_N,P_N)$ describing the current FRF estimate and the information accumulated from previously selected measurements, without requiring storage of the complete measurement history.
These quantities are used next to assess the informativeness of newly acquired measurements and determine whether they should be transmitted.

\subsection{Event-Triggering Rule}
\label{subsec:event_trigger}
We make explicit the dependence of the number of selected measurements on the current candidate sampling instant \(t_k\) denoting by $N(t_k):=\sum_{i=1}^{\infty}\mathbbm{1}_{\{\tau_i<t_k\}}$ the number of event-selected measurements incorporated into the estimator before $t_k$.
Immediately before observing $y(t_k)$, the sensing unit has access to the current estimate $\hat{\theta}_{N(t_k)}$, the matrix $P_{N(t_k)}$, and the regressor $\varphi(t_k)$, which is known from the excitation signal.
We collect these quantities in 
\begin{equation}
    \mathcal I(t_k)
    :=
    \left\{
        \varphi(t_k),
        \hat{\theta}_{N(t_k)},
        P_{N(t_k)}
    \right\}.
    \label{eq:information_state}
\end{equation}
The triggering problem is to determine the additional information that the new measurement $y(t_k)$ provides relative to the pre-measurement information state $\mathcal I(t_k)$.

To assess the relevance of a newly acquired measurement, we evaluate it relative to the predictions of the current model.
Measurements that are well explained by the current model provide little additional information, whereas observations associated with high predictive uncertainty or inconsistent with the model prediction are more informative.
These two effects can be characterized through the predictive density, whose negative logarithm provides a measure of predictive surprise.
In particular, measurements lying in regions of low predictive density provide strong evidence that the model should be updated. 
We first establish this interpretation under nominal conditions, namely when the FRF parameter vector remains constant over the measurements represented
by the current RLS state and no forgetting is applied ($\lambda=1$).
This setting, formalized in Lemma \ref{lem:predictive}, provides an exact statistical interpretation of the quantities used subsequently in the event trigger.
Their role in the piecewise-LTI setting with exponential forgetting is discussed after the lemma.

\begin{lemma}[Predictive surprise]
\label{lem:predictive}
Consider the recursive least-squares \eqref{eq:RLS_multisine} with $\lambda=1$, and suppose that the FRF parameter vector is constant over the measurements incorporated into the recursion, and that the corresponding measurements are collected in steady state.
Under Assumption~\ref{assump:noise}, the negative log predictive density of a candidate measurement $y(t_k)$, conditioned on the pre-measurement information state $\mathcal I(t_k)$, satisfies
\begin{align}
    &-\log p\!\left(y(t_k)\mid \mathcal I(t_k)\right) = \nonumber \\ 
    & \frac12\log \left(1+\varphi^{\hop}(t_k)P_{N(t_k)}\varphi(t_k)\right)+\frac12\frac{\varepsilon^2(t_k)}{S_k}+\frac12\log(2\pi\sigma^2),
    \nonumber
\end{align}
where $S_k = \sigma^2 \left(1+\varphi^{\hop}(t_k)P_{N(t_k)}\varphi(t_k)\right)$, and the first term is  constant.
\end{lemma}
\vspace{-0.2cm}

\begin{proof}
Under the state assumptions, the regression error $\varepsilon(t_k)$ reduces to the measurement noise, so that $y(t_k) = \varphi^\hop(t_k)\theta + e(t_k)$.
For $\lambda=1$, the RLS \eqref{eq:RLS_multisine} admits the standard linear-Gaussian interpretation.
In particular, with $\hat{\theta}_0= 0_{2M+1}$ and $P_0=\alpha I_{2M+1}$, the corresponding parameter prior distribution is $\theta \sim \mathcal N \left(0_{2M+1}, \sigma^2\alpha I_{2M+1}\right)$.
After $N(t_k)$ selected measurements have been incorporated, the uncertainty represented by the RLS state is therefore $\theta \mid \mathcal I(t_k) \sim \mathcal N \left(\hat{\theta}_{N(t_k)}, \sigma^2P_{N(t_k)} \right)$.
Under Assumption~\ref{assump:noise}, $e(t_k)\sim\mathcal N(0,\sigma^2)$, so that the measurement noise is independent of both $\theta$ and $\mathcal I(t_k)$.
Since $y(t_k)=\varphi^{\hop}(t_k)\theta+e(t_k)$ is an affine transformation of the jointly Gaussian random variables $\theta$ and $e(t_k)$, the predictive distribution $p(y(t_k)\mid\mathcal I(t_k))$ is Gaussian.
Its conditional mean is
\[
\begin{aligned}
\mu_k\coloneqq&\,\mathbb E \!\left[y(t_k) \mid \mathcal I(t_k) \right] \\ =
&\,\varphi^\hop(t_k) \mathbb E\!\left[\theta \mid \mathcal I(t_k) \right] + \mathbb E\!\left[ e(t_k) \mid \mathcal I(t_k) \right] \\
= &\,\varphi^{\hop}(t_k) \hat{\theta}_{N(t_k)},
\end{aligned}
\]
where the last equality follows since $e(t_k)$ is  independent of $\mathcal I(t_k)$, and $\mathbb E[e(t_k)]=0$.
Its conditional variance is
\[
\begin{aligned}
S_k
\coloneqq &\, \operatorname{Var}\!\left(y(t_k) \mid \mathcal I(t_k) \right) \\
= &\,\operatorname{Var} \!\left(\varphi^{\hop}(t_k)\theta \mid \mathcal I(t_k) \right) + \operatorname{Var}\left(e(t_k)\mid \mathcal I(t_k)\right) \\
= &\, \sigma^2 \left(1+\varphi^{\hop}(t_k)P_{N(t_k)} \varphi(t_k) \right),
\end{aligned}
\]
where the cross-term vanishes because $e(t_k)$ is conditionally independent of $\theta$ given $\mathcal I(t_k)$ and we use the identity $\operatorname{Var}(A\,\theta)=A\operatorname{Var}(\theta)A^\hop$ for deterministic matrices $A$.

Evaluating the corresponding Gaussian density at the newly acquired
measurement gives
\[
    p\bigl(y(t_k)\mid\mathcal I(t_k)\bigr)
    =
    \frac{1}{\sqrt{2\pi S_k}}
    \exp\!\left(-\frac{(y(t_k)-\mu_k)^2}{2S_k}\right),
\]
which yields the result in the lemma.
\end{proof}

Lemma~\ref{lem:predictive} decomposes the predictive surprise of a candidate measurement into two complementary terms.
The first, $\log\!\left(1+\varphi^\hop(t_k) P_{N(t_k)} \varphi(t_k) \right)$, depends only on the information state available before the candidate output is observed and quantifies the potential uncertainty reduction in the direction associated with $\varphi(t_k)$.
The second term, $\varepsilon^2(t_k)/S_k$, depends on the realized output and quantifies how unexpected the observation is relative to the current model prediction.
A large value indicates model mismatch.
Hence, a measurement can be informative either because it excites a poorly known direction of the current FRF estimate, or because the observed output is inconsistent with the current model.

The practical setting considered in this work departs from the conditions of Lemma~\ref{lem:predictive} in two respects: the system is piecewise LTI according to Assumption \ref{ass:abrupt_fault}, and exponential forgetting with $\lambda<1$ is introduced to allow the local estimator to adapt.
Immediately after a change in the system dynamics, the RLS estimate generally contains information associated with both the previous and the current dynamics.
As new event-selected measurements are incorporated, however, the influence of older measurements is progressively attenuated. 
Therefore, the effective information retained by the estimator becomes increasingly dominated by measurements generated under the current LTI regime.
In this sense, exponential forgetting progressively restores the setting underlying Lemma \ref{lem:predictive}.

This comes at the cost that, for $\lambda<1$, $\sigma^2P_N$ no longer has the exact covariance interpretation used in Lemma \ref{lem:predictive}.
Nevertheless, $R_N=P_N^{-1}$ remains the exponentially weighted information matrix associated with the event-selected regressors, as shown in \eqref{eq:P_inverse}.
Consequently, \(\varphi^\hop(t_k)P_{N(t_k)}\varphi(t_k)\) still quantifies the inverse information available along the candidate regressor direction.
The residual term remains meaningful for a complementary reason.
Following a change in the system dynamics, the current estimate may temporarily represent the previous dynamics, causing new measurements to deviate systematically from predictions.
Therefore, $\varepsilon^2(t_k)/S_k$ provides a normalized measure of this model mismatch and promotes the selection of measurements that contain evidence of changed dynamics.

These observations motivate an event-triggering rule that retains the uncertainty-mismatch structure revealed by Lemma~\ref{lem:predictive}, with two practical modifications.
For the uncertainty component, since the logarithm is strictly increasing, we use $U_k := \varphi^{\hop}(t_k)P_{N(t_k)}\varphi(t_k)$ instead of $\log(1+\varphi^{\hop}(t_k)P_{N(t_k)}\varphi(t_k))$.
This preserves the ordering induced by the uncertainty component while providing a simpler quantity for trigger design.
For the mismatch component, directly using the instantaneous normalized innovation squared $\varepsilon^2(t_k)/S_k$ may make the trigger sensitive to isolated noise-induced outliers.
We therefore introduce the exponentially weighted moving average
\begin{equation}
    z_k
    =
    \beta z_{k-1}
    +(1-\beta)\frac{\varepsilon^2(t_k)}{S_k},
    \qquad \beta\in[0,1),
    \label{eq:EWMA}
\end{equation}
so that evidence of model mismatch is accumulated over successive measurements rather than inferred from a single innovation.
The recursion \eqref{eq:EWMA} is initialized with $z_0=1$, matching the expected value of the normalized innovation squared under the nominal conditions of Lemma~\ref{lem:predictive}.

\begin{remark}
The parameter $\beta$ determines the trade-off between responsiveness to changes in the dynamics and sensitivity to isolated noise fluctuations.
For small $\beta$, $z_k$ reacts rapidly to new prediction errors, whereas larger values of $\beta$ provide stronger averaging and require sustained model mismatch before significantly affecting the trigger.
\end{remark}

Combining the two components gives the event-trigger
\begin{equation}
    \Gamma\!\left(\mathcal I(t_k),y(t_k)\right)
    = U_k + \gamma z_k,
    \label{eq:event_trigger}
\end{equation}
where $\gamma>0$ controls the trade-off between exploration of uncertain directions and reaction to model mismatch.
The measurement $y(t_k)$ is selected and transmitted whenever $\Gamma\!\left(\mathcal I(t_k),y(t_k)\right)\geq\eta$.
The information properties of \eqref{eq:event_trigger} are analyzed in Section \ref{sec:analysis}.

\section{Information Properties of Event-Triggered Data Selection}
\label{sec:analysis}

This section analyzes the event-triggering mechanism from an information perspective.
First, we establish a direct relation between the uncertainty component $U_k$ and the incremental D-optimal information gain associated with including a candidate measurement in the identification dataset.
Then, we show that, under persistent excitation of the candidate regressors, the event trigger prevents the information retained from the selected measurements to become arbitrarily small.

We consider the information matrix $R_{N(t_k)}$ associated with the ${N(t_k)}$ previously selected measurements.
To quantify the overall information available for identification, we consider the D-optimal criterion $\log\det R_{N(t_k)}$, which accounts jointly for the information accumulated along all parameter directions.
If the candidate measurement at $t_k$ is selected, the information matrix after the RLS update satisfies $R_{{N(t_k)}+1} = \lambda R_{N(t_k)} + \varphi(t_k)\varphi^\hop(t_k)$, where $\lambda R_{N(t_k)}$ represents the information retained from the previously selected measurements after forgetting.
To compare the post-update information matrix $R_{{N(t_k)}+1}$ with this discounted pre-update information, we define the incremental D-optimal information gain as
\begin{equation}
    \Delta_D(t_k) := \log\det R_{{N(t_k)}+1} - \log\det(\lambda R_{N(t_k)}).
    \label{eq:incremental_doptimal}
\end{equation}
Proposition \ref{prop:incremental_doptimal} establishes a direct relation between $\Delta_D(t_k)$ and the uncertainty component $U_k$.

\begin{proposition}[Incremental D-optimal information gain]
\label{prop:incremental_doptimal}
For the RLS \eqref{eq:RLS_multisine}, the incremental D-optimal information gain associated with retaining the candidate measurement at $t_k$ is
\begin{equation}
    \Delta_D(t_k) = \log\left(
        1+\frac{1}{\lambda}
        \varphi^\hop(t_k)P_{N(t_k)}\varphi(t_k)
    \right).
    \label{eq:doptimal_gain}
\end{equation}
\end{proposition}

\begin{proof}
Since $R_{N(t_k)}\succ0$ and $\lambda>0$, the matrix $\lambda R_{N(t_k)}$ is nonsingular.
Applying the matrix determinant lemma
\[
    \det(A+uv^\hop)
    =
    \det(A)\left(1+v^\hop A^{-1}u\right)
\]
with $A=\lambda R_{N(t_k)}$ and $u=v=\varphi(t_k)$ gives
\begin{align*}
    \det R_{{N(t_k)}+1}=
    \det(\lambda R_{N(t_k)})
    \left(
        1+\frac{1}{\lambda}
        \varphi^\hop(t_k)
        P_{N(t_k)}
        \varphi(t_k)
    \right).
\end{align*}
Taking logarithms and subtracting $\log\det(\lambda R_{N(t_k)})$ gives \eqref{eq:doptimal_gain}.
\end{proof}

\vspace{-0.4cm}

Proposition~\ref{prop:incremental_doptimal} gives the uncertainty component a direct interpretation in experiment design.
Since the mapping $U_k \mapsto \log\left(1+{U_k}/{\lambda}\right)$ is strictly increasing, $U_k$ ranks candidate measurements in exactly the same order as their incremental D-optimal information gain.
A large value of $U_k$ therefore identifies a candidate whose inclusion would produce a large increase in the log-determinant of the retained information matrix.
This interpretation establishes a direct connection with the offline sampling-design framework of \cite{dozzi2026d}.
There, sampling instants are chosen before the experiment to maximize the information content of the resulting identification data.
Here, the same D-optimal principle is applied sequentially to evaluate the marginal information contribution of each newly acquired measurement before deciding whether it should be retained.
The uncertainty term $U_k$ can therefore be viewed as an online, event-driven counterpart of the D-optimal sampling criterion used in the offline design.

This result characterizes the information contribution of an individual candidate measurement.
A further question is whether repeated rejection of measurements can cause the retained information to vanish along some parameter direction. Lemma \ref{lem:bounded_uncertainty} shows that, under persistent excitation of the candidate regressors, the retained information matrix remains uniformly positive definite.

\begin{lemma}[Bounded information]
\label{lem:bounded_uncertainty}
Consider the recursive least-squares \eqref{eq:RLS_multisine}, with measurements selected according to the event-triggering rule \eqref{eq:event_trigger}.
Assume that the candidate regressor sequence is uniformly persistently exciting, i.e., there exist an integer $W\geq1$ and $\mu>0$ such that
\begin{equation}
    \sum_{j=k}^{k+W-1}
    \varphi(t_j)\varphi^\hop(t_j)
    \succeq
    \mu I_{2M+1},
    \qquad \forall k \in \mathbb{N}.
    \label{eq:candidate_PE}
\end{equation}
For every $k$, after processing $W$ candidate measurements $t_k,\ldots,t_{k+W-1}$, the retained information matrix satisfies
\begin{equation}
    R_{N(t_{k+W})}
    \succeq
    \frac{\lambda^W\mu}
         {1+W\eta}I_{2M+1}.
    \label{eq:R_uniform_bound}
\end{equation}
\end{lemma}

\begin{proof}
Let $\delta_k$ indicate whether the candidate measurement acquired at $t_k$ is selected, i.e., $\delta_k=1$ if $\Gamma_k \geq \eta$ and $\delta_k=0$ if $\Gamma_k < \eta$ .
First, consider a candidate measurement $y(t_k)$ that is not selected.
This implies $U_k < {\eta}$ since $\Gamma_k<\eta$ and $\gamma z_k\geq0$.
Between two consecutive candidate instants,
\begin{equation}
    R_{N(t_{k+1})}
    \succeq
    \lambda R_{N(t_k)}
    +
    \delta_k
    \varphi(t_k)\varphi^\hop(t_k).
    \label{eq:candidate_information_update}
\end{equation}
Indeed, equality holds when $\delta_k=1$, whereas for $\delta_k=0$, $R_{N(t_{k+1})} = R_{N(t_k)} \succeq \lambda R_{N(t_k)}$.
Fix an arbitrary unit vector $v$, with $\|v\|_2=1$.
For notational convenience, define $r_{j}=v^\hop R_{N(t_{j})}v$ and $\phi_j=v^\hop \varphi(t_j)$.
Premultiplying and postmultiplying \eqref{eq:candidate_information_update} by $v^\hop$ and $v$ gives
\begin{equation}
    r_{k+1} \geq \lambda r_k + \delta_k   \left|\phi_k\right|^2.
    \label{eq:rj_update}
\end{equation}
Now define $c := \frac{\mu}{1+W\eta}$.
We distinguish two cases.

\emph{Case 1:}
Suppose there exists $\ell\in\{k,\ldots,k+W\}$ such that $r_\ell \geq c$.
Since \eqref{eq:rj_update} implies $r_{j+1}\geq\lambda r_j$, $r_{k+W} \geq \lambda^{k+W-\ell}\,r_\ell \geq \lambda^W c$.

\emph{Case 2:}
Suppose that $r_{j}<c$ for every $j=k,\ldots,k+W$.
For every rejected candidate $t_j$, the Cauchy--Schwarz inequality yields
\begin{align}
    \left| \phi_j \right|^2=\left|v^\hop\varphi(t_j)\right|^2
    &=
    \left|
        \big(R_{N(t_j)}^{1/2}v\big)^\hop
        \big(R_{N(t_j)}^{-1/2}\varphi(t_j)\big)
    \right|^2
    \leq r_{j} U_j.
    \nonumber
\end{align}
Since $r_{j}<c$ and $U_j<\eta$, every rejected candidate $t_j$ satisfies $\left|\phi_j\right|^2 < c\eta$.
Let $\mathcal S_k$ and $\mathcal R_k$ denote the selected and rejected candidates in the considered window, respectively,
\begin{align*}
    \mathcal S_k &:= \left\{j\in\{k,\ldots,k+W- 1\}:\delta_j=1 \right\}, \\
    \mathcal R_k &:= \left\{j\in\{k,\ldots,k+W-1\}:\delta_j=0 \right\}.
\end{align*}
Since $\sum_{j=k}^{k+W-1}\left|\phi_j\right|^2 \geq \mu$ from \eqref{eq:candidate_PE}, we have
\begin{align}
    \sum_{j\in\mathcal S_k}
    \left|\phi_j\right|^2
    &=
    \sum_{j=k}^{k+W-1}
    \left|\phi_j\right|^2
    -
    \sum_{j\in\mathcal R_k}
    \left|\phi_j\right|^2
    \nonumber>
    \mu-W\eta c.
    \label{eq:selected_information_lower}
\end{align}
By definition $\mu-W\eta c=c$, thus $\sum_{j\in\mathcal S_k} \left|\phi_j\right|^2 >c$.
Iterating \eqref{eq:rj_update} over the $W$ candidate instants gives
\begin{align}
    r_{k+W}
    &\geq
    \lambda^W  r_{k}
    +
    \sum_{j=k}^{k+W-1}
    \lambda^{k+W-1-j}
    \delta_j
    \left|\phi_j\right|^2
    \nonumber \geq
    \lambda^W c.
    \nonumber
\end{align}
Thus, in both cases, $r_{k+W} \geq \lambda^W c$.
Since $v$ is an arbitrary unit vector, the statement follows.
\end{proof}

Lemma \ref{lem:bounded_uncertainty} provides a global information-retention property of the event-triggering mechanism, showing that event-based data selection does not lead to an unbounded loss of information.
Despite discarding measurements and exponentially forgetting previously selected data, the event-triggered data selection guarantees that the retained information matrix remains uniformly positive definite under persistent excitation conditions.
The mismatch component can result in additional selections, but is not required for this information-retention guarantee.
Since $P_{N(t_k)}=R_{N(t_k)}^{-1}$, the result implies that the uncertainty associated with the selected data cannot grow arbitrarily large, i.e.,
\begin{equation}
    P_{N(t_k+W)}
    \preceq
    \frac{1+W\eta}
         {\lambda^W\mu}I_{2M+1}.
\end{equation}
For the multisine regressor \eqref{eq:varphi}, under non-aliasing of the sampled excitation frequencies and nonzero excitation amplitudes, \eqref{eq:candidate_PE} holds with $W=2M+1$ \citep{gonzalez2025finite}.
The constant $\mu>0$ is given by the minimum eigenvalue of the Gram matrix $\sum_{j=k}^{k+2M}\varphi(t_j)\varphi^\hop(t_j)$ and is independent of the starting index $k$.

\section{Simulation Results}
\label{sec:simulation_results}

The proposed event-triggered data-selection framework is evaluated on a mass-spring-damper-mass system, with the first mass connected to ground through an additional spring-damper.
A force is applied to the first mass and the displacement of the second mass is measured.
The mechanical parameters and identification settings are summarized in Table \ref{tab:simulation_parameters}.
The system is initialized at rest, and at time $T_1=5000$ s the coupling stiffness abruptly changes from $k_{12,i}$ to $k_{12,f}$.
The transients are not explicitly modeled by the estimator and are included in the regression error in \eqref{eq:linear_regression}.
The measurement noise is zero-mean Gaussian with variance $\sigma^2=0.5$, in accordance with Assumption~\ref{assump:noise}.
The system is excited by a multisine input with frequencies $\Omega=\{0.5, 1, 1.5, 2, 2.5, 3, 3.5, 4, 5, 7,  8, 10, 12\}$, and its FRF is estimated online at the same frequencies.
The amplitudes and phases are drawn independently and uniformly from $[5,6.5]$ and $[0,2\pi]$, respectively, with the same realization used throughout all simulations.
Measurements are acquired periodically at the local unit, while only those satisfying \eqref{eq:event_trigger} are transmitted to the central unit.

\begin{table}[t]
    \centering
    \caption{Simulation parameters.}
    \label{tab:simulation_parameters}
    \begin{tabular}{lll}
        \toprule
        Parameter & Symbol & Value \\
        \midrule
        \multicolumn{3}{l}{\textit{System parameters}} \\
        Mass 1, Mass 2                  & $m_1$, $m_2$       & $1$ kg \\
        Coupling damping coefficient     & $c_{12}$         & $0.1$ Ns/m\\
        Ground damping coefficient     & $c_{ground}$         & $0.2$ Ns/m\\
        Initial coupling stiffness       & $k_{12,i}$       & $12.5$ N/m \\
        Changed coupling stiffness       & $k_{12,f}$       & $7.0$ N/m \\
        Ground stiffness & $k_{ground}$ & $2.0$ N/m \\
        \midrule
        \multicolumn{3}{l}{\textit{Identification and triggering parameters}} \\
        Candidate sampling period & $T_s$    & $0.1$ s \\
        Forgetting factor          & $\lambda$  & 0.995 \\
        EWMA parameter             & $\beta$    & 0.9 \\
        Trigger threshold          & $\eta$     & 0.4 \\
        Mismatch weight            & $\gamma$ & 0.2 \\
        \bottomrule
    \end{tabular}
\end{table}

To assess the information retained by the selected measurements, the central unit applies the same RLS estimator \eqref{eq:RLS_multisine} to the received
data.
This choice is made for evaluation purposes, so that differences in estimation performance can be attributed to the measurements made available to the central unit rather than to differences in the identification algorithm.
In a practical implementation, however, the central unit does not need to employ the same recursive estimator.
The transmitted measurements can be processed by any identification method based on the same FRF regression model.
Identification performance is quantified through the FRF estimation error $e_\theta(t_k) := \left\|        \theta(t_k)-\hat{\theta}_{N(t_k)} \right\|_2$, where $\theta(t_k)$ contains the true FRF values at the excited frequencies at time $t_k$ and $\hat{\theta}_{N(t_k)}$ denotes the current estimate at the central unit.
Since exponential forgetting is employed, the estimator retains a finite effective amount of past information and the estimation error is therefore not expected to vanish asymptotically \citep{goel2020recursive}.
Instead, the simulations assess both the steady-state estimation accuracy and the ability of the transmitted data to support adaptation following the change in dynamics.

\begin{figure}
    \centering
    \includegraphics[width=0.99\linewidth]{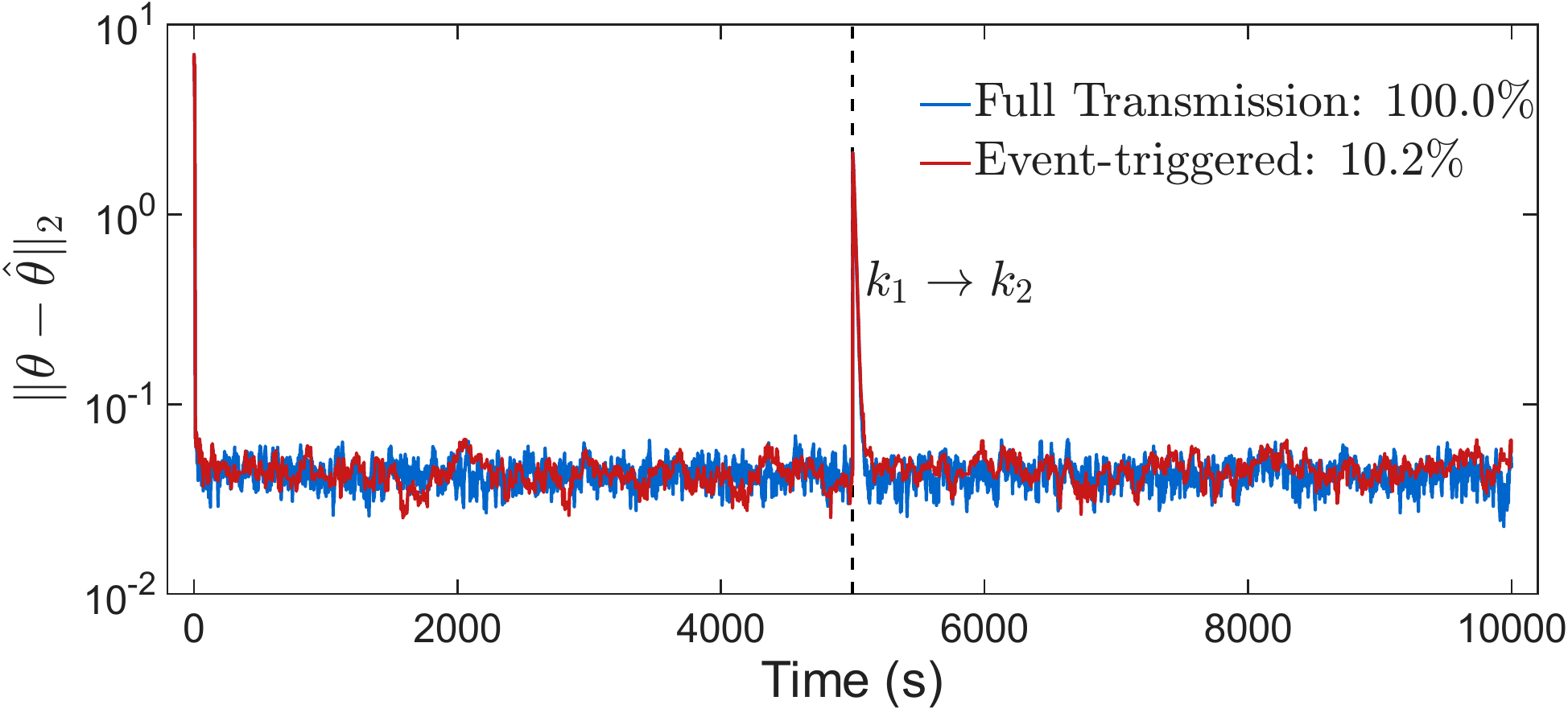}
    \caption{FRF estimation error under full periodic transmission and event-triggered data selection.}
    \label{fig:full_comparison}
\end{figure}

We first compare the proposed event-triggered data-selection scheme with full periodic transmission, where all measurements acquired by the local unit are transmitted to the central estimator.
The simulation consists of $N_s=100\,000$ candidate sampling instants with sampling period $T_s=0.1$~s.
The event-triggered scheme selects approximately $10\,000$ measurements, corresponding to a communication ratio of about $10\%$.
Figure~\ref{fig:full_comparison} shows that the event-triggered estimator achieves essentially the same FRF estimation performance as full transmission, both during initial convergence and after the abrupt change in dynamics.
At initialization, little information about the FRF is available, and thus measurements are selected at a high rate. 
Similarly, following the change in dynamics, the increased model mismatch causes the trigger to temporarily select measurements more frequently, allowing the estimator to adapt at approximately the same rate as when all measurements are available.
Once the estimate has converged to the current dynamics, the event rate decreases and only measurements deemed sufficiently informative are transmitted.
Thus, comparable steady-state identification accuracy and adaptation to changes are obtained while reducing the number of transmitted measurements by approximately $90\%$.

\begin{figure}
    \centering
    \includegraphics[width=0.99\linewidth]{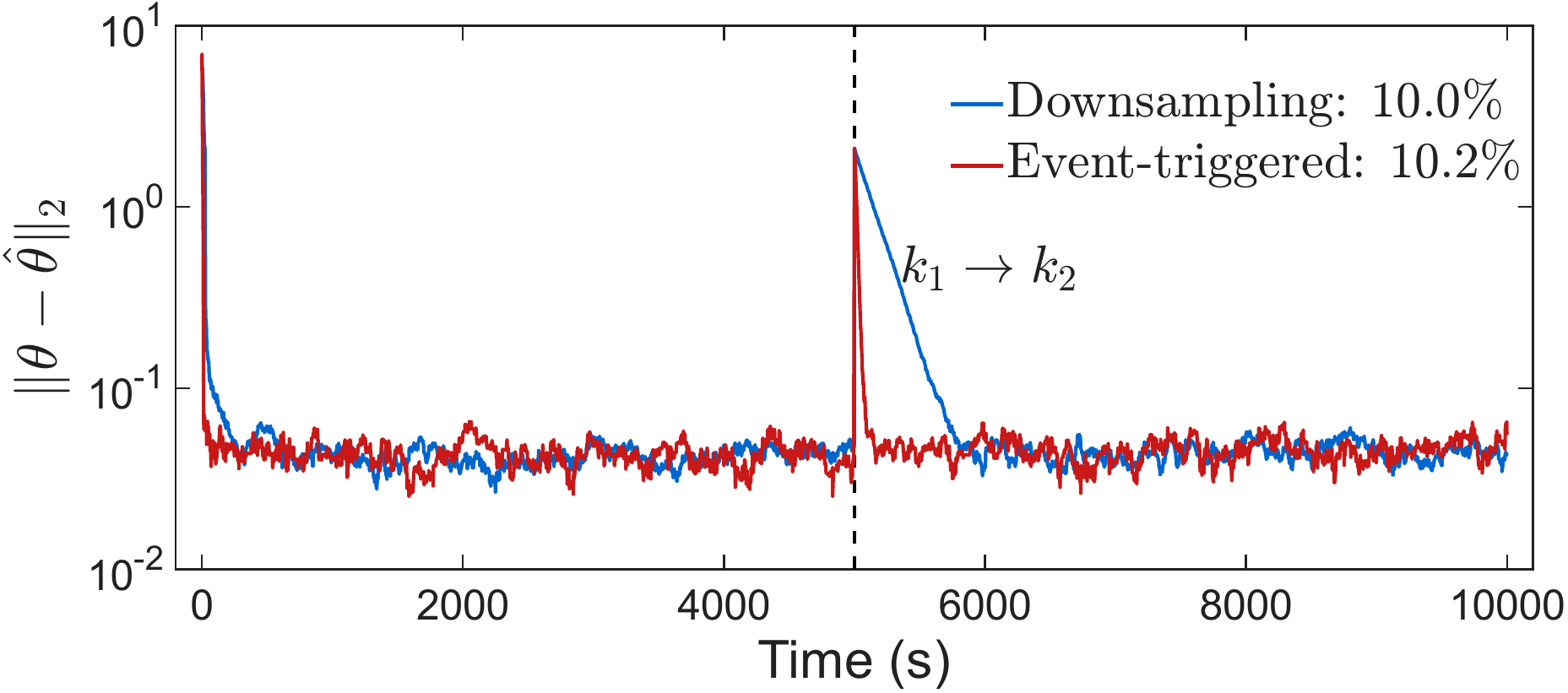}
    \caption{FRF estimation error under uniform downsampling and event-triggered data selection.}
    \label{fig:downsampling_comparison}
\end{figure}

However, a comparable reduction in communication could be achieved more simply by uniformly downsampling the periodically acquired measurements.
We therefore compare the proposed approach with uniform downsampling under the same communication budget.
Since the event-triggered scheme transmits approximately $10\%$ of the candidate measurements, the periodic baseline retains one out of every ten samples, corresponding to a transmission interval of $1$~s.
Unlike the proposed approach, this strategy selects measurements independently of their relevance for identification: the transmission instants are fixed a priori and cannot adapt to the information content of the data.

\begin{figure}
    \centering
    \includegraphics[width=0.99\linewidth]{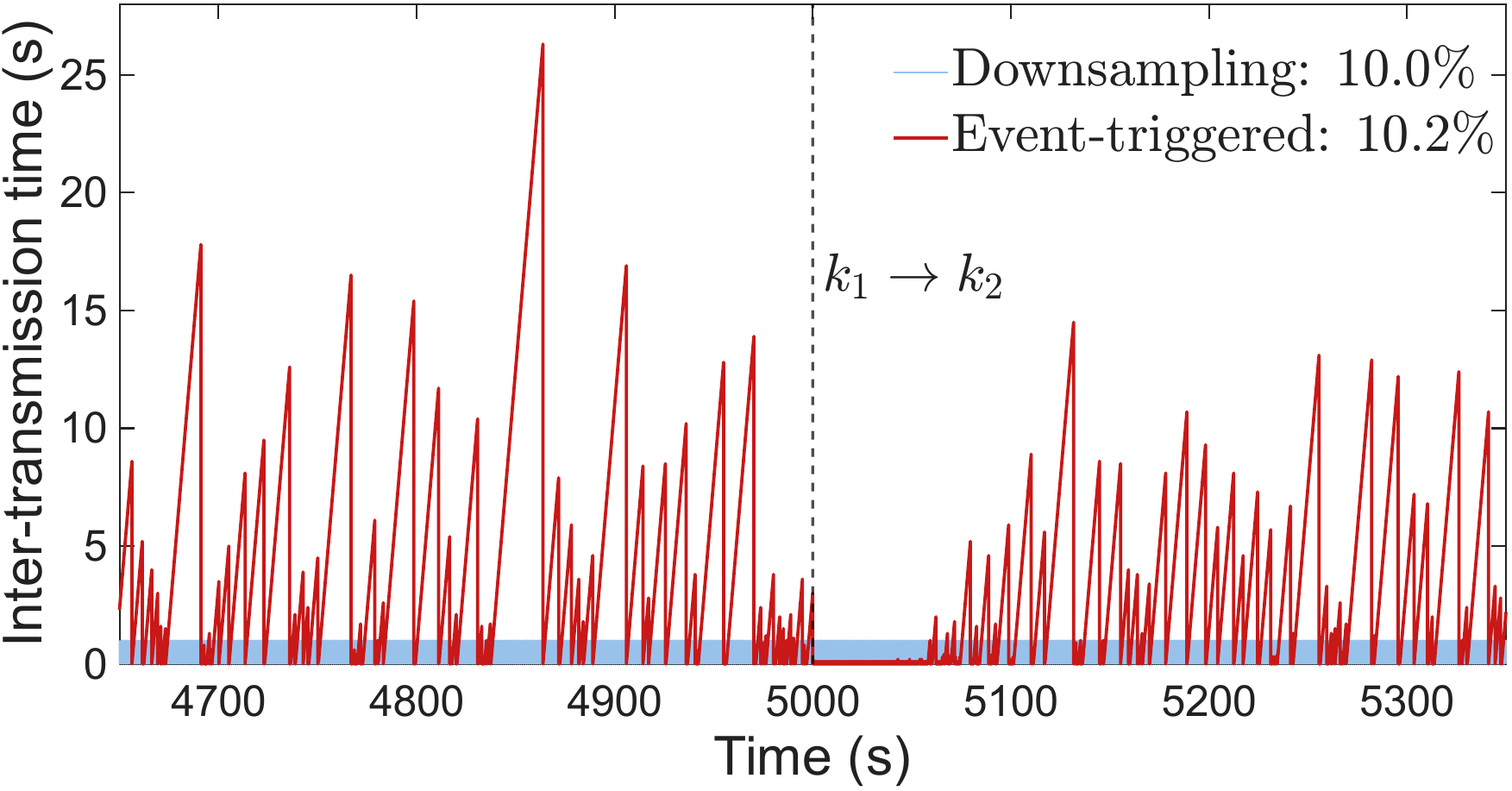}
    \caption{Inter-event transmission times around the abrupt change in system dynamics.}
    \label{fig:inter_event_time}
\end{figure}

Figure \ref{fig:downsampling_comparison} shows that both strategies attain a comparable estimation error once the estimator has converged.
Following the abrupt change in the system dynamics, however, the event-triggered approach adapts substantially faster by temporarily increasing the transmission rate in response to the increased mismatch between the observed measurements and the estimates.
As shown in Figure~\ref{fig:inter_event_time}, the inter-event time decreases to the candidate sampling period $T_s$ immediately after the change, indicating that essentially every available measurement is selected.
As the estimator adapts to the new dynamics, the transmission rate decreases again.
Therefore, the event trigger distributes a comparable communication budget in time, allocating transmissions to periods in which newly acquired measurements are most relevant for identification and adaptation.

\section{Conclusion}
\label{sec:conclusions}
This paper presented an event-triggered data-selection approach for online frequency-domain identification in distributed sensing architectures.
Periodically acquired measurements are evaluated locally and only relevant measurements are transmitted for remote identification.
The triggering rule combines an uncertainty component, promoting measurements that contribute new information about the FRF, with a model-mismatch component that provides responsiveness to changes in the system dynamics.
The uncertainty component is shown to rank candidate measurements according to their incremental D-optimal information gain and to guarantee that the retained information matrix remains uniformly positive definite despite data selection and exponential forgetting.
Simulation results showed that the method can substantially reduce communication while maintaining estimation accuracy, and fast adaptation to abrupt changes.

\bibliography{references}
\end{document}